\documentclass[preprint,aps,prb,showpacs,showkeys,preprintnumbers,superscriptaddress,floatfix]{revtex4-1}
\usepackage{graphicx}
\usepackage{dcolumn}
\usepackage{float}
\usepackage{bm}
\usepackage{latexsym}
\usepackage{array}
\usepackage{amssymb}
\usepackage{amsfonts}
\usepackage{amsmath}
\usepackage{color}
\newcommand{\comment}[1]{}
\newcommand{\done}[1]{}

\newcommand{\dagga}{\dagger}
\newcommand{\be}{\begin{equation}}
\newcommand{\ee}{\end{equation}}
\newcommand{\bea}{\begin{eqnarray}}
\newcommand{\eea}{\end{eqnarray}}
\newcommand{\baa}{\begin{align}}
\newcommand{\eaa}{\end{align}}

\newcommand{\ket}[1]{|#1\,\rangle}
\newcommand{\bra}[1]{\langle\,#1|}

\usepackage{ifthen}
\usepackage{calc}
\usepackage{url}
\usepackage{fancyvrb}
\usepackage[utf8]{inputenc}
\usepackage[english]{babel}%

\usepackage{mathrsfs}
\usepackage{comment}
\usepackage[font=small,labelfont=bf,tableposition=top]{caption}

\begin{document}

\title{A finite field approach to solving the Bethe Salpeter equation}
%%%%%%%%%%%%%%%%%%%%%%%%%%%%%%%%%%%%%%%%%%%%%%%%%%%%%%%%%%%%%%%%%%%%%%%%%%%%%%%
\author{Ngoc Linh Nguyen} 
\affiliation{Institute for Molecular Engineering, The University of Chicago, Chicago, Illinois 60637, USA}
\author{He Ma} 
\affiliation{Institute for Molecular Engineering, The University of Chicago, Chicago, Illinois 60637, USA}
\affiliation{Department of Chemistry, The University of Chicago, Chicago, Illinois 60637, USA}
\author{Marco Govoni}
\affiliation{Institute for Molecular Engineering, The University of Chicago, Chicago, Illinois 60637, USA}
\affiliation{Materials Science Division and Institute for Molecular Engineering, Argonne National Laboratory, Lemont, Illinois 60439, USA}
\author{Francois Gygi}
\affiliation{Department of Computer Science, University of California Davis, Davis, California 95616, USA}
\author{Giulia Galli}
\affiliation{Institute for Molecular Engineering, The University of Chicago, Chicago, Illinois 60637, USA}
\affiliation{Department of Chemistry, The University of Chicago, Chicago, Illinois 60637, USA}
\affiliation{Materials Science Division and Institute for Molecular Engineering, Argonne National Laboratory, Lemont, Illinois 60439, USA}

\date{\today}% It is always \today, today,

\begin{abstract}
We present a method to compute optical spectra and exciton binding energies of molecules and solids based on the solution of the Bethe-Salpeter equation (BSE) and the calculation of the screened Coulomb interaction in finite field. The method does not require the explicit evaluation of  dielectric matrices nor of virtual electronic states, and can be easily applied without resorting to the random phase approximation. In addition, it utilizes localized orbitals obtained from Bloch states using bisection techniques, thus greatly reducing the complexity of the calculation and enabling the efficient use of  hybrid functionals to obtain  single particle wavefunctions. We report exciton binding energies of several molecules and absorption spectra of condensed systems of unprecedented size, including water and ice samples with hundreds of atoms.
\end{abstract}
%%%%%%%%%%%%%%%%%%%%%%%%%%%%%%%%%%%%%%%%%%%%%%%%%%%%%%%%%%%%%%%%%%%%%%%%%%%
\maketitle

The ability to simulate optical properties of materials from first principles is key to building predictive strategies for the design of new materials and molecules, as well as to interpreting increasingly complex experimental results~\cite{doi:10.1002/aenm.201400915, Petousis2017, pham_modelling_2017}.
The last three decades have witnessed a tremendous success of many-body perturbation theory (MBPT)~\cite{RevModPhys.74.601,martin_reining_ceperley_2016} in the description of the interaction of molecules and condensed matter with light. MBPT, a Green's function method, can be used to accurately compute various excitation properties, based on single particle energies and orbitals obtained, e.g. within density functional theory (DFT)~\cite{PhysRev.136.B864,PhysRev.140.A1133}.
In particular, by solving the Dyson equation~\cite{Fetter} within the GW approximation~\cite{Hedin_PR_1965} and the Bethe-Salpeter equation (BSE)~\cite{RevModPhys.74.601,PhysRev.84.1232}, one can accurately predict the energy of charged and neutral  excitations~\cite{RevModPhys.74.601}, excitonic and charge transfer states~\cite{doi:10.1063/1.3655352, Galli_jcp_2010}, and optical absorption spectra~\cite{PhysRevLett.43.387,PhysRevLett.49.1519,PhysRevLett.80.3320,Blase_chem_soc_rev_2018,Galli_chem_soc_rev_2013}.
However, the solution of the BSE is computationally demanding, more so, for example, than the use of  
time-dependent density functional theory (TD-DFT) with semi-local or hybrid exchange-correlation (xc) functionals~\cite{PhysRevLett.52.997, Marques2003,Baroni_prl_2006}. Therefore, TD-DFT is still widely used to compute absorption spectra, albeit often yielding less accurate results than the BSE.

The unfavorable cost of conventional approaches~\cite{PhysRevLett.80.4510,PhysRevB.62.4927,PhysRevLett.80.4514,MARINI20091392,MARTINSAMOS20091416,Casida_book_2011} to solve the BSE is mainly due to the evaluation of explicit summations over virtual states and to the need of evaluating and inverting large dielectric matrices.
In particular, the straightforward diagonalization of the two-body exciton Hamiltonian in the basis of electron-hole pairs 
requires a workload of order $O(N^6)$, where $N$ is the number of electrons in the system~\cite{PhysRevLett.80.4510,PhysRevLett.80.3320}. 
A formulation of the BSE without empty states that sidesteps the diagonalization of the two-body exciton Hamiltonian, and does not require the inversion of dielectric matrices was recently proposed~\cite{Galli_jcp_2010, PhysRevB.85.045116,Galli_chem_soc_rev_2013}, and shown to accurately yield absorption spectra over a wide range of frequencies using the Liouville-Lanczos algorithm~\cite{Baroni_prl_2006,Baroni_jcp_2008}. 
A distinctive feature of this formalism based on density matrix perturbation theory (DMPT) is the utilization of projective dielectric eigenpotentials (PDEP)~\cite{PhysRevB.79.245106,PhysRevB.78.113303} to compute screened exchange integrals. 
Despite the advantages of the DMPT formulation and its more favourable $O(N^4$) scaling, drawbacks remain, including the need to extrapolate the results as a function of the number of dielectric eigenpotentials  and, most importantly,  the difficulty to use  DFT calculations with hybrid density functionals~\cite{doi:10.1063/1.472933,PhysRevX.6.041002,PhysRevB.89.195112} as a starting point for the BSE solution. 
   
In this Letter, we present a novel method to solve the BSE by performing calculations in finite electric fields. The two key features of the method are: (i)  the direct evaluation of the screened Coulomb interaction in finite field (FF), thus eliminating the need to compute  dielectric matrices altogether; (ii)  the use of a compact, localized representation~\cite{PhysRevLett.102.166406} of the ground state Kohn-Sham (KS) wavefunctions, leading to a great reduction of the cost to evaluate screened exchange integrals. We show that these features lead to a major improvement in the efficiency of the BSE solution and, importantly, to the straightforward use of the results of hybrid functionals as a starting point for GW and BSE calculations. The FF-BSE can be used to compute not only the properties of single molecules or solids, but its solution may be easily coupled to first principles molecular dynamics (FPMD) simulations to obtain, e.g. optical spectra over multiple snapshots extracted from trajectories at finite temperature and pressure, as we show below. We report examples for the optical spectra of liquid water and ice as obtained by averaging over multiple trajectories, for systems with up to 2,048 electrons. In addition, we present the results of calculations using ground state wavefunctions computed with hybrid functionals~\cite{PhysRevB.89.195112}.

\textit{Method}. Absorption spectra of solids and molecules can be obtained by computing  the imaginary part of the macroscopic dielectric function $\text{Im}\epsilon^M_{ij}=4\pi Im \frac{\partial P_i}{\partial E_j}$, where $\mathbf{E}$ is the macroscopic electric field, and $\mathbf{P}=-\frac{1}{\Omega}\text{Tr}\left\{ \hat{\mathbf{r}}\hat{\rho}\right\}$ the macroscopic polarization, $\hat{\mathbf{r}}$ is the position operator, and $\hat{\rho}$  the density matrix.\cite{note_units} We obtain  $\frac{\partial P_i}{\partial E_j}$ from the solution of the Liouville equation for the density matrix.\cite{Baroni_prl_2006} For a system described by a mean-field Hamiltonian $\hat{H}(\hat{\rho})$ subject to a monochromatic electrostatic potential $\phi(\omega)=-\mathbf{E}(\omega)\cdot\mathbf{r}$, the time evolution of the density matrix is given by the Liouville equation $ \omega \hat{\rho} = [\hat{H}(\hat{\rho})-\hat{\phi}, \hat{\rho}]$. Upon linearization, we obtain the first order variation of the density matrix as the solution of the following non-homogeneous linear system:
\begin{equation}\label{eig_lou_eq_1} 
    (\omega - \mathcal{L})
    \Delta\hat{\rho}
    = - [\hat{\phi}, \hat{\rho}^{o}]\,,
\end{equation}
where $\hat{\rho}^{o}$ is the unperturbed density matrix. The Liouville superoperator $\mathcal{L}$ acting on $\Delta\hat{\rho}$ is defined as
\begin{equation}\label{lou_qe_2}
    \mathcal{L}\Delta\hat{\rho} = [\hat H^{o}, \Delta\hat{\rho}] + [\Delta \hat{V}_{\rm H}, \hat{\rho}^{o}] + [\Delta\hat{\Sigma}, \hat{\rho}^{o}] \,,
\end{equation}
where $\hat H^{o}$ is the unperturbed Hamiltonian, and $\Delta\hat{V}_H$ and $\Delta\hat{\Sigma}$ are the first-order variation of the Hartree and the exchange-correlation (xc) self-energy induced by $\Delta\hat{\rho}$, respectively.
The change in polarization induced by $\mathbf{E}$, entering the definition of the absorption spectrum, can hence be expressed as $
\frac{\partial P_i}{\partial E_j} =-\frac{1}{\Omega} \text{Tr}\left\{\hat{r}_i 
\frac{\partial \Delta \hat{\rho}}{\partial E_j}
\right\}$.
As previously noted~\cite{Galli_jcp_2010}, the homogeneous linear system corresponding to Eq.~\ref{eig_lou_eq_1} is a secular equation with neutral excitation energies as eigenvalues; these energies are equivalent to those obtained by solving the BSE with static screening if an effective Hamiltonian and the COHSEX self-energy~\cite{Hedin_PR_1965, martin_reining_ceperley_2016} are utilized for $\hat H^{o}$ and $\Delta\hat{\Sigma}$, respectively.
However, unlike BSE solvers based on the diagonalization of the two-particle electron-hole Hamiltonian\cite{RevModPhys.74.601}, Eq.~\ref{eig_lou_eq_1} can be solved without defining a transition space, and hence a direct product of occupied and unoccupied active subspaces. In order to avoid such definition and the need to compute virtual electronic orbitals, we introduce the auxiliary functions $\ket{a^j_v}=\hat{P}_c\frac{\partial \Delta \hat{\rho}}{\partial E_j}\ket{\varphi_v}$, where $\varphi_v$ is the $v$-th occupied state of the unperturbed Hamiltonian (with energy $\varepsilon_v$); $\hat{P}_{c}=1-\sum_{v=1}^{N_{\text{occ}}}\left|\varphi_v\right\rangle\left\langle\varphi_v\right|$ is the projector onto the unoccupied manifold~\cite{Baroni_RevModPhys.73.515}, and $N_{\text{occ}}$ is the number of occupied states. It has been shown that an Hermitian solution of Eq.~\ref{eig_lou_eq_1} can be written as $\frac{\partial \Delta \hat{\rho}}{\partial E_j}=\sum_{v=1}^{N_\text{occ}}\left( \ket{\varphi_v}\bra{a^j_v}+ \ket{a^j_v}\bra{\varphi_v}\right)$, and the functions $a^j_v$ are obtained from the solution of the following non-homogeneous linear systems~\cite{suppl_material}:
\begin{equation}\label{eq:linearsys}
    \sum_{v^\prime=1}^{N_\text{occ}} \left(\omega\delta_{vv^\prime} -D_{vv^\prime}-K^{1e}_{vv^\prime}+K^{1d}_{vv^\prime}\right)\ket{a^j_{v^\prime}} = \hat{P}_c \hat{r}_j\ket{\varphi_v}\,,
\end{equation}
where the three terms on the RHS of Eq.~\ref{lou_qe_2} are:
\begin{eqnarray}
    D_{vv^\prime} \ket{a^j_{v^\prime}} &=& \hat{P}_c\left(\hat{H}^{o}-\varepsilon_v\right)\delta_{vv^\prime} \ket{a^j_{v^\prime}}\,, \\
    K^{1e}_{vv^\prime} \ket{a^j_{v^\prime}} &=& 2 \hat{P}_c \left(\int d\mathbf{r^\prime}v_c(\mathbf{r},\mathbf{r}^\prime)\varphi^\ast_{v^\prime}(\mathbf{r}^\prime)a^j_{v^\prime}(\mathbf{r}^\prime)\right) \varphi_v(\mathbf{r})\,, \label{eq:k1e} \\
    K^{1d}_{vv^\prime} \ket{a^j_{v^\prime}} &=&  \hat{P}_c \tau_{vv^\prime}(\mathbf{r})a^j_{v^\prime}(\mathbf{r})\,. \label{eq:k1d}
\end{eqnarray}
and we have defined the  screened integrals $\tau_{vv^\prime}(\mathbf{r}) = \int{{W(\mathbf{r},\mathbf{r^\prime})}\varphi_{v}(\mathbf{r^\prime}) \varphi_{v'}^{*}(\mathbf{r^\prime}) d\mathbf{r^\prime}}$, where $W$ and $v_c$ are the screened and bare Coulomb interactions, respectively. Eq.~\ref{eq:linearsys} can be solved for multiple frequencies using the Lanczos algorithm\cite{note_broadening}.  The evaluation of the integrals $\tau_{vv^\prime}$ represents the most expensive part of the calculation because it entails a computation of the dielectric matrix. Recently, Eq.~\ref{eq:linearsys}-\ref{eq:k1d} were solved using Kohn-Sham (KS) states as input, using  DFT calculations with semi-local functionals, and a spectral representation of the dielectric matrix via its eigenvectors, called  projective dielectric eigenpotentials (PDEP).\cite{PhysRevB.85.045116,PhysRevB.85.035316,PhysRevB.87.165203} 

\color{black}
Here we introduce a new approach with two key features: (i) the screened integrals are directly computed from finite field calculations avoiding any explicit evaluation of the dielectric matrix; in addition, (ii) the total number of required integrals, in principle equal to $N_\text{occ}^2$, is  reduced to a much smaller number that scales linearly with the system size, by using a compact, localized representation of single particle wavefunctions. The very same representation is adopted to increase the efficiency of hybrid-DFT calculations\cite{PhysRevLett.102.166406}, leading to a formulation of BSE which requires the very same workload when using local or hybrid-DFT starting points. We now illustrate steps (i) and (ii) in detail.

Using the definition of the screened Coulomb interaction in terms of the density-density response function, $W=v_c+v_c\chi v_c$, we express the screened integrals as $\tau_{vv^\prime}=\tau^u_{vv^\prime}+v_c \chi \tau^u_{vv^\prime}$, where  $\tau^u_{vv^\prime}(\mathbf{r})=\int{{v_c(\mathbf{r},\mathbf{r^\prime})}\varphi_{v}(\mathbf{r^\prime}) \varphi_{v'}^{*}(\mathbf{r^\prime})d\mathbf{r^\prime}}$ are obtained by multiplying orbitals in real space and then applying the bare Coulomb potential $v_c$ in reciprocal space.
For each $\tau^u_{vv^\prime}$ we  determined two densities ($\rho^\pm_{vv^\prime}$) by solving self-consistently the uncoupled-perturbed KS equations with Hamiltonian $(\hat H^{o}\pm \tau^u_{vv^\prime})$. The screened exchange integrals are then obtained as: 
\begin{equation}
\tau_{vv^\prime}(\mathbf{r}) =  \tau^u_{vv^\prime}(\mathbf{r}) + \int{{v_c(\mathbf{r},\mathbf{r^\prime})} \frac{\rho_{vv^\prime}^+(\mathbf{r^\prime}) - \rho_{vv^\prime}^-(\mathbf{r^\prime})}{2} d\mathbf{r^\prime}} \,,
\label{tauutotau}
\end{equation}
where a central finite difference formula was used to compute the linear variation of the density, i.e. $\chi \tau^u_{vv^\prime}$.  The algorithm described above was implemented by coupling  the WEST~\cite{doi:10.1021/ct500958p} and Qbox~\cite{5388642} codes, operating in client-server mode~\cite{note_amplitude}, thus enabling massive parallel calculations by assigning independent finite field calculations to different Qbox instances, which may be started at any point during, e.g. a first principle MD simulation.

Next we reduced the number of integrals to compute, in principle equal to $N_{\text{occ}}^2$, by localizing single particle wavefunctions in appropriate regions of real space and neglecting those orbital pairs that do not overlap. To do so we used the recursive bisection technique\cite{PhysRevLett.102.166406}, whereby orbitals are truncated in subdomains of variable size while controlling the 2-norm error caused by the truncation procedure. This technique was previously used to improve the efficiency of calculations of exact exchange integrals and is here applied to {\it screened} exchange integrals. 
When using orbital bisection, a unitary transformation $U$: $|\tilde{\varphi}_m\rangle = \sum_v U_{mv}|\varphi_v\rangle$ of the occupied KS states is evaluated,  and used to transform the   matrix $\tilde{\tau} = U \tau U^\dagga $ (and similarly $\tau^u$) into a sparse form, where only a relatively small number of selected elements need to be computed using Eq.~\ref{tauutotau}. The number of required non-zero screened integrals scales linearly with system size. Different types of localized orbitals were used previously to solve the BSE, e.g. atomic-orbital basis sets~\cite{PhysRevB.92.075422}, or maximally localized Wannier orbitals~\cite{RevModPhys.84.1419,PhysRevB.95.075415}. However, there are several advantages of the localization technique used here: (i) it is adaptive, i.e. the orbitals  can be localized in domains of different shapes and sizes; (ii) it  allows to systematically control the localization
error with a single parameter, and (iii) it is consistently applied to reduce the number of screened integrals and, at the same time, to speed up hybrid-DFT calculations\cite{doi:10.1021/ct3007088,doi:10.1021/acs.jctc.5b00826}. Hence the workload of our calculations is of $O(N^4)$ for the evaluation of $\tau_{vv^\prime}$ and of $O(N^3)$ for the evaluation of Eq.~\ref{eq:k1d}, {\it irrespective of whether semilocal or hybrid functionals are used}.  This is an important achievement, especially for the study of optical properties of materials, e.g. complex oxides, for which semi-local functionals do not even represent a qualitatively correct starting point to solve the BSE. In addition, we note that the computational gain of the method presented here increases as the size of the system increases, that is the prefactor in our calculations is increasingly smaller, compared to that of density functional perturbation theory calculations, as the size of the system increases (see  SI).

  \begin{figure}
    \centering
    %\begin{minipage}{0.52\textwidth}
    %\includegraphics[width=\textwidth]{ver.png}
    %\end{minipage}
    %\begin{minipage}{0.45\textwidth}
    %\begin{tabular}[b]{c|ccc}\hline
    %  (eV) & $\chi^{RPA}$@PBE & $\chi$@PBE & $\chi$@PBE0 \\ \hline
    %  MAD & 0.95 & 0.86 & 0.31 \\
    %  RMSD & 1.06 & 0.98 & 0.59 \\\hline
    %\end{tabular}
    \includegraphics[width=0.5\textwidth]{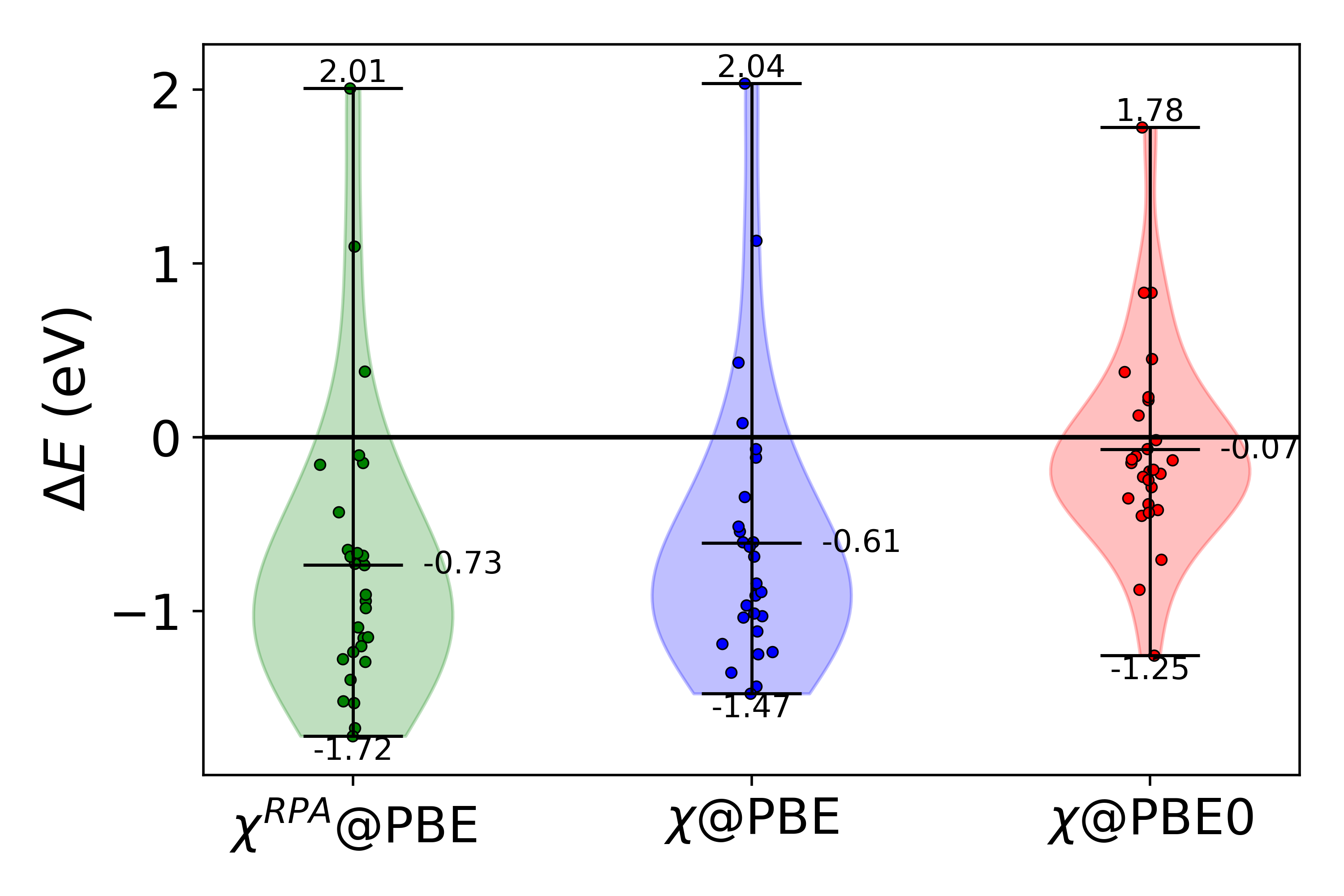}
    %\end{minipage}
    \caption{\label{thiel_set_plot} 
    The lowest singlet excitation energies of the 28 molecules of the Thiel's set computed by
solving the  Bethe Salpeter equation in finite field (FF-BSE) with (green) and without (blue) the
Random Phase Approximation (RPA), using the PBE and the PBE0 hybrid
functional (red). Results are compared ($\Delta E$) with the best theory estimates obtained using quantum chemistry methods~\cite{doi:10.1080/00268970903549047,doi:10.1063/1.3499598}. The horizontal lines denote the maximum, mean, and minimum of the distribution of results, compared with quantum chemistry methods. $\chi$ denotes the response function computed with and without the RPA. The numerical values are reported in the
SI~\cite{suppl_material}.}
  \end{figure}

%%%%%%%%%%%%%%%%%%%%%%%%%%%%%%%%%%%%%%%%%%%%%%%%%%%%%%%%%%%%%%%%%%%%%%%%%%%%%%%%

\textit{Verification and Validation}. To demonstrate the accuracy of the FF-BSE methodology, we first calculated the neutral singlet excitation energies for the Thiel's set~\cite{doi:10.1080/00268970903549047,doi:10.1063/1.3499598}, which consists of 28 small organic molecules. We compared our results with the best theoretical estimates as obtained from quantum chemistry calculations, i.e. coupled cluster and complete active space second-order perturbation theory using the aug-cc-pVTZ atomic basis set~\cite{doi:10.1080/00268970903549047,doi:10.1063/1.3499598}. This molecular set was recently used to benchmark GW-BSE~\cite{doi:10.1063/1.4922489, doi:10.1063/1.4983126, doi:10.1021/acs.jctc.5b00304} and TD-DFT calculations (with PBE0~\cite{doi:10.1021/acs.jctc.5b00304} and dielectric-dependent hybrid functionals~\cite{PhysRevX.6.041002}).
We evaluated the screened integrals in Eq.~\ref{tauutotau} with and without the RPA, and using either the Perdew–Burke–Ernzerhof (PBE)~\cite{PhysRevLett.77.3865} or the PBE0 hybrid functional~\cite{doi:10.1063/1.478522}. 
As shown in Fig.\ref{thiel_set_plot}, we obtained a good agreement with benchmark calculations, thus validating our methodology for molecules. A small change is observed when we compute the screened integrals with and without the Random Phase Approximation (RPA). Our results also show that BSE calculations based on G$_0$W$_0$ starting from the PBE (PBE0) ground state underestimate excitation energies by $\sim 0.7$ eV ($\sim 0.1$ eV). The improvement observed with the PBE0 functional underscores the importance of an accurate ground state starting point. 
We also validated our method for solid LiF and compared our calculations with experiment and previous results (see SI).

%%%%%%%%%%%%%%%%%%%%%%%%%%%%%%%%%%%%%%%%%%%%%%%%%%%%%%%%%%%%%%%%%%%%%%%%%%%%%%%%
\begin{figure} 
   \includegraphics[width=0.5\textwidth]{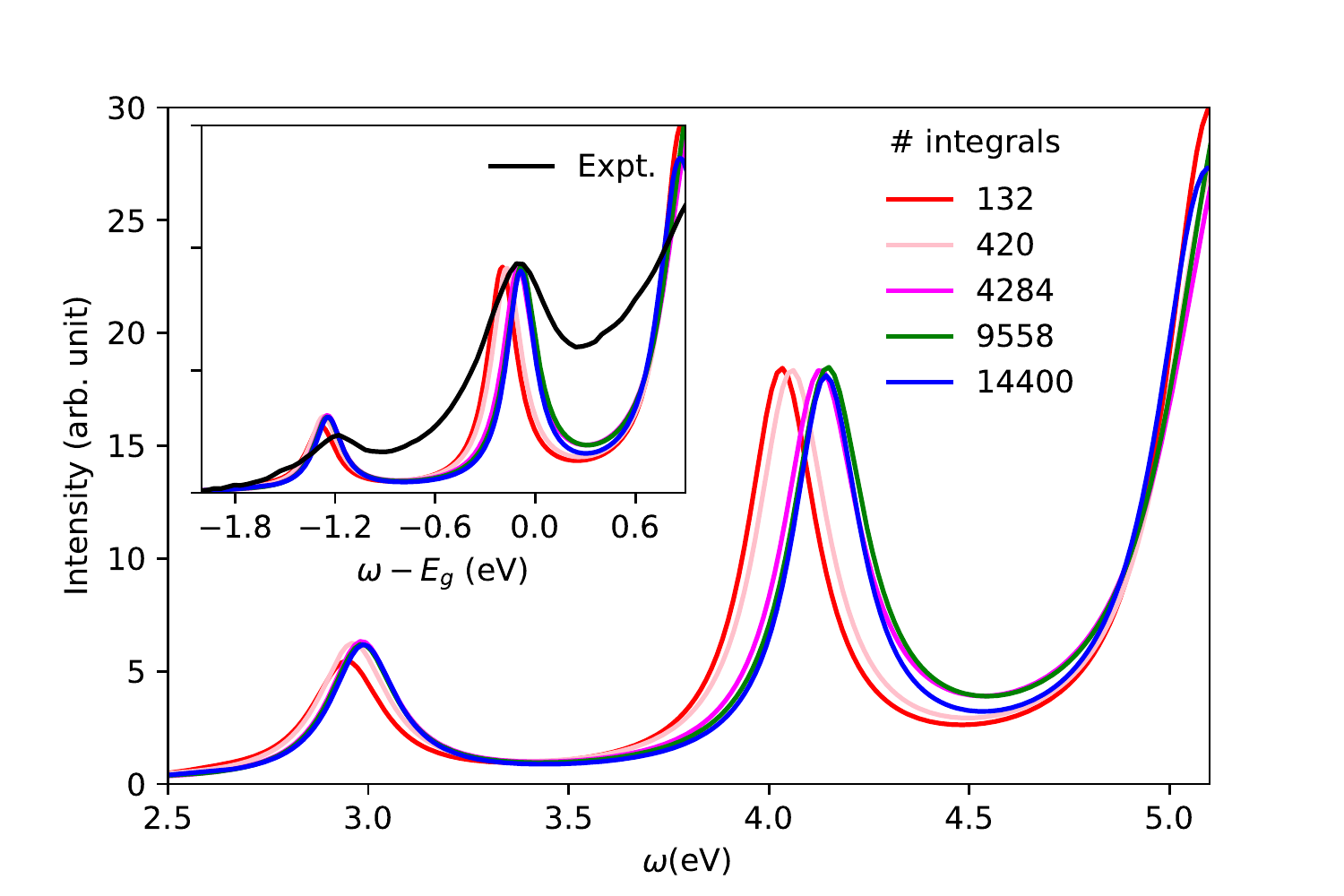}
\caption{\label{abs_spec_vs_ovl} Optical absorption spectra of C$_{60}$ in the gas phase computed by solving the BSE with several thresholds $\xi$  for the screened exchange integrals. The resulting number of integrals is indicated.  The inset shows the same spectra plotted as a function of $\omega-E_{\rm g}$, and compared with experiment~\cite{1347-4065-32-11B-L1667}. $E_{\rm g}$ is the electronic gap. Note that an accurate spectrum is obtained when using 4,284 integrals instead of the total number which is more than three times larger (14,400).}
\end{figure}
%%%%%%%%%%%%%%%%%%%%%%%%%%%%%%%%%%%%%%%%%%%%%%%%%%%%%%%%%%%%%%%%%%%%%%%%%%%%%%%%

Next, we show how the use of bisected orbitals can reduce the computational cost of BSE calculations of  optical spectra of the C$_{60}$ fullerene in the gas phase. The computed electronic gap at the optimized PBE geometry and at the G$_0$W$_0$@PBE level of theory is 4.23 eV. This value is smaller than that  obtained at the experimental geometry (4.55 eV) at the same level of theory, consistent with  Ref.~\onlinecite{PhysRevB.91.245105}, and it is $\sim$~0.7~eV lower than the experimental value, estimated as the energy difference between the measured ionization potential and the electronic affinity~\cite{doi:10.1063/1.459778, doi:10.1063/1.478732}. To evaluate the exciton binding energy, E$_{\rm b}^{\rm ex}$, we computed the energy difference between the electronic gap and the lowest optically-allowed singlet excited state (the lowest neutral eigenstate has $T^{1}_{\rm g}$ symmetry).
C$_{60}$ has 120 doubly occupied valence states, and in principle 14,400 integrals should be evaluated. As shown in Fig.~\ref{abs_spec_vs_ovl}, the number of screened integrals entering Eq. \ref{tauutotau} can be greatly reduced without hardly any loss for accuracy in the computed absorption spectrum.

%%%%%%%%%%%%%%%%%%%%%%%%%%%%%%%%%%%%%%%%%%%%%%%%%%%%%%%%%%%%%%%%%%%%%%%%%%%%%%%%
\begin{figure*} 
   \includegraphics[width=0.7\textwidth]{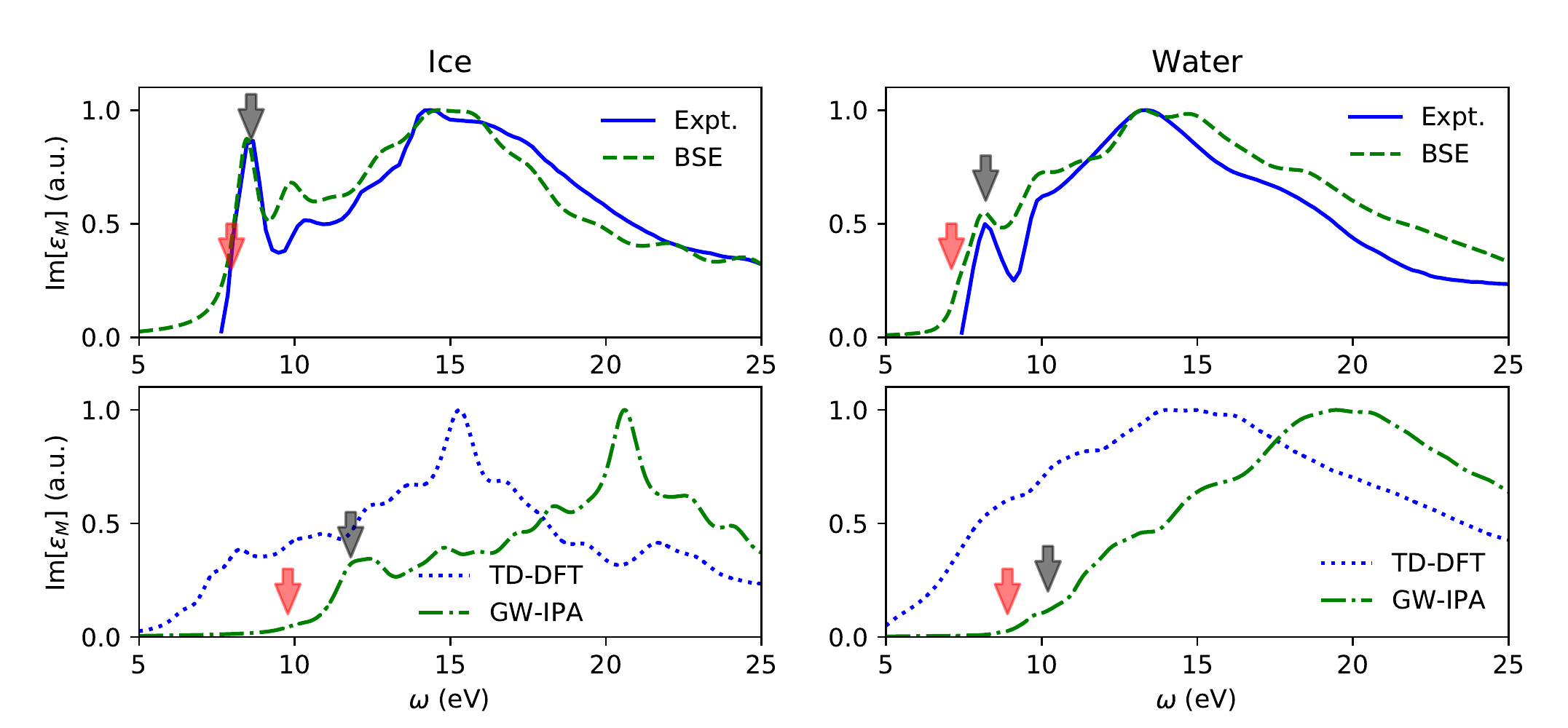}
   \caption{\label{abs_spec_water_ice} Imaginary part of the macroscopic dielectric constant ($\epsilon_M$) as a function of the photon frequency ($\omega$) for a proton-disorder hexagonal ice model (left panel) and liquid water (right panel)
   computed as an average over nine samples extracted from path-integral molecular dynamics (PIMD) trajectories~\cite{Gaiduk2018} generated with the MBPol potential. Experimental results (from Refs.~\onlinecite{doi:10.1063/1.1681563}  and~\onlinecite{doi:10.1021/j100244a065} for water and ice, respectively) are shown by the blue solid lines.
   The black and red arrows indicate the positions of the first excitonic peak and the onset of the spectra, respectively.} 
\end{figure*}
%%%%%%%%%%%%%%%%%%%%%%%%%%%%%%%%%%%%%%%%%%%%%%%%%%%%%%%%%%%%%%%%%%%%%%%%%%%%%%%%

\textit{Results}. In the last part of this Letter, we report results for the optical absorption spectra of liquid water and ice.
Even though the first measurement of these spectra dates back to 1974~\cite{doi:10.1021/j100244a065,doi:10.1063/1.1681563,PhysRevA.5.2523},  experimental estimates of the exciton binding energy, E$_{\rm b}^{\rm ex}$, are yet uncertain due to  uncertainties in the values of  
ice and water electronic gaps~\cite{doi:10.1021/jacs.6b00225,Gaiduk2018} and to the presence of a low energy tail in the absorption spectra ($\sim1.0$ eV)~\cite{doi:10.1021/j100244a065,doi:10.1063/1.1681563,PhysRevA.5.2523} hampering a precise determination of the onset energy.
Thus far, only a few GW-BSE computations of the optical spectra of water and ice have been carried out; several theoretical studies used rather small unit cells ($\simeq 17$ water molecules) and approximations for the static dielectric matrix \cite{doi:10.1063/1.4929468,PhysRevLett.100.207403,PhysRevLett.106.187403,PhysRevLett.94.037404,PhysRevLett.97.137402} (e.g. homogeneous electron gas model). 
Here, we performed calculations for several samples of 64 water molecules of liquid water, extracted from MD trajectories~\cite{Gaiduk2018} and for 96 water molecules of a proton-disorder hexagonal ice model~\cite{doi:10.1021/acs.jpclett.5b00901}, 
whose structure was optimized with the PBE0 functional at 0 K~\cite{doi:10.1021/acs.jpclett.5b00901}. 
In Fig.\ref{abs_spec_water_ice}, we  compare our results with experiments~\cite{doi:10.1021/j100244a065,doi:10.1063/1.1681563}. 
Due to the underestimation of the G$_0$W$_0$@PBE~\cite{PhysRevB.89.060202, PhysRevLett.117.186401,doi:10.1021/jacs.6b00225} electronic gaps of both systems, GW-BSE absorption spectra are red-shifted with respect to the experimental one. 
Hence we aligned the first peak of the computed GW-BSE spectrum with experiment, and we shifted the TD-DFT and GW-IPA (independent particle) spectra by the same energy. 
We found a remarkable agreement between GW-BSE and experiment both for the relative energy positions and intensities of the peaks over a wide range of energy. 
As expected, the TD-DFT and GW-IPA approximations predict significantly different spectra. We examined the influence of the DFT wavefunctions and eigenvalues chosen as starting point of the calculation, finding a good qualitative agreement between spectra for one water configuration computed at the DFT-PBE and dielectric hybrid [dielectric-dependent hybrid (DDH)\cite{PhysRevB.89.195112}] level of theory (see Fig.~5 of SI\cite{suppl_material}). 
 We also investigated the effect of different structural models on the computed spectra, by comparing results 
obtained using trajectories generated with the MB-pol potential ~\cite{doi:10.1021/ct400863t} and path integral MD, with those computed for PBE trajectories, extracted from the PBE400 set~\cite{doi:10.1063/1.5018116}.
Our results show a broadening of the averaged PIMD absorption spectrum, and a red-shift of $\sim 0.5$ eV with respect to the averaged FPMD@PBE spectrum [see Fig.~7 of SI\cite{suppl_material}]. 

The exciton binding energies of  liquid water and ice were computed using 64 and 96 water molecules, and were evaluated as the energy differences between the onset (E$^{\rm ex}_{\rm b1}$) and the first main peak (E$^{\rm ex}_{\rm b2}$) of the absorption spectra (marked by black and red arrows respectively in Fig.~\ref{abs_spec_water_ice}).  We obtained E$^{\rm ex}_{\rm b1} = 1.64$ eV and 1.82 eV, and E$^{\rm ex}_{\rm b2} = 2.3$ eV and 3.12 eV for water and ice, respectively. The values for E$^{\rm ex}_{\rm b2}$ are consistent with those reported in previous calculations, i.e. 2.5~\cite{PhysRevLett.97.137402} and 3.2~\cite{PhysRevLett.94.037404} eV. Finally, we also performed calculations for a larger supercell including 256 water molecules (2048 valence electrons) and concluded that size effects, although not fully negligible, are rather minor on the value of the exciton binding energies (of the order of $\sim$0.2-0.3 eV).

\textit{Conclusions}. We have presented a novel method to solve the BSE in finite field, which not only avoids the calculations of virtual electronic states, but avoids all together the calculation of dielectric matrices. In addition, our formulation uses linear combinations of Bloch orbitals that are localized in appropriate regions of real space, leading to substantial computational savings. There are several advantages of the method presented here: calculations beyond the RPA are straightforward and
the complexity and scaling of solving the BSE is {\it the same} when using local or hybrid-DFT starting points. As a consequence, the method proposed here leads to an improvement in both accuracy and efficiency in the calculations of optical spectra of large molecular and condensed systems, and to the ability of coupling such computations with first principles molecular dynamics. 

\begin{acknowledgments}
We gratefully acknowledge Sijia Dong and Huihuo Zheng for helpful conversations. This work was supported by MICCoM, as part of the Computational Materials Sciences Program funded by the U.S. Department of Energy, Office of Science, Basic Energy Sciences, Materials Sciences and Engineering Division. N.L.N. is supported by the Swiss National Science Foundation (grant no. P300P2\_171422). This research used resources of the Argonne Leadership Computing Facility, which is a DOE Office of Science User Facility supported under contract DE-AC02-06CH11357.
\end{acknowledgments}

\bibliography{biblio}% Produces the bibliography via BibTeX.

\begin{thebibliography}{10}
\bibitem{note_units}
Atomic units are used. $\Omega$ is the volume of the unit cell. Absorption
  spectra of molecules are obtained defining the macroscopic polarization as
  $\mathbf{P}=-\text{Tr}\left\{\hat{\mathbf{r}}\hat{\rho}\right\}$.
\bibitem{note_broadening}
All computed spectra have been obtained with a Lorentzian broadening of 0.1 eV.
\end{thebibliography}
\bibliographystyle{aip}

\end{document}